**Speaker Identification in the Shouted Environment Using Suprasegmental Hidden Markov Models**

Ismail Shahin

Electrical and Computer Engineering Department

University of Sharjah

P. O. Box  27272

Sharjah, United Arab Emirates

Tel: (971) 6 5050967, Fax: (971) 6 5050877

E-mail: ismail@sharjah.ac.ae

# Abstract

In this paper, Suprasegmental Hidden Markov Models (SPHMMs) have been used to enhance the recognition performance of text-dependent speaker identification in the shouted environment. Our speech database consists of two databases: our collected database and the Speech Under Simulated and Actual Stress (SUSAS) database. Our results show that SPHMMs significantly enhance speaker identification performance compared to Second-Order Circular Hidden Markov Models (CHMM2s) in the shouted environment. Using our collected database, speaker identification performance in this environment is 68% and 75% based on CHMM2s and SPHMMs respectively. Using the SUSAS database, speaker identification performance in the same environment is 71% and 79% based on CHMM2s and SPHMMs respectively.

**Keywords:** hidden Markov models; second-order circular hidden Markov models; shouted environment; speaker identification; suprasegemental hidden Markov models.

## 1. Introduction

Speaker recognition is the process of automatically recognizing who is speaking on the basis of individuality information in speech waves. Speaker recognition systems come in two flavors: speaker identification systems and speaker authentication (verification) systems.

Speaker identification is the process of determining from which of the registered speakers a given utterance comes. Speaker identification systems can be used in criminal investigations to determine the suspected persons produced the voice recorded at the scene of the crime [1]. Speaker identification systems can also be used in civil cases or for the media. These cases include calls to radio stations, local or other government authorities, insurance companies, or recorded conversations, and many other applications.

Speaker authentication is the process of determining whether a speaker corresponds to a particular known voice or to some other unknown voice. The applications of speaker authentication systems involve the use of voice as a key to confirm the identity claim of a speaker. Such services include banking transactions using a telephone network, database access services, security control for confidential information areas, remote access to computers, tracking speakers in a conversation or broadcast, and many other areas.

Speaker recognition systems typically operate in one of two cases: text-dependent (fixed text) case or text-independent (free-text) case. In the text-dependent case, utterances of the same text are used for both training and testing (recognition). On the other hand, in the text-independent case, training and testing involve utterances from different texts.

The process of speaker recognition can be divided into two categories: "open set" and "closed set". In the "open set" category, a reference model for the unknown speaker may not exist; whereas, in the "closed set" category, a reference model for the unknown speaker should be available to the system.

## 2. Motivation

The majority of researchers who work in the areas of speech recognition and speaker recognition focus their work on speech under the neutral talking condition and the minority of the researchers focus their work on speech under the stressful talking conditions. The neutral talking condition is defined as the talking condition in which speech is produced assuming that speakers are in a "quiet room" with no task obligations. The Stressful talking conditions can be defined as the talking conditions that cause speakers to vary their production of speech from the neutral talking condition.

Some talking conditions are designed to simulate speech produced by different speakers under real stressful talking conditions. Hansen, Cummings, Clements, Bou-Ghazale, Zhou, and Kaiser used SUSAS database in which eight talking conditions are used to simulate speech produced under real stressful talking conditions and three real talking conditions [2], [3], [4]. The eight talking conditions are: neutral, loud, soft, angry, fast, slow, clear, and question. The three real talking conditions are: 50% task (cond50), 70% task (cond70), and Lombard. The 50% task and the 70% task comprise utterances recorded from subjects engaged in tracking tasks under different levels of workload (the level of workload is higher in 70% than that in 50%). The Lombard effect occurs when speakers vary their speech characteristics in order to increase intelligibility when speaking in a noisy environment. Chen used six talking conditions to simulate speech under real stressful talking conditions [5]. These conditions are: neutral, fast, loud, Lombard, soft, and shouted.

Very few researchers who focus their work on speech under stressful talking conditions consider studying speech under the shouted talking condition [5], [6], [7], [8]. Therefore, the number of publications in the areas of speech recognition and speaker recognition under this talking condition is very limited. The shouted talking condition can be defined as when speakers shout, their intention is to produce a very loud acoustic signal, either to increase its range (distance) of transmission or its ratio to background noise.

Speaker identification systems under the shouted talking condition can be used in the applications of talking condition identification systems. Such systems can be used in medical applications where computerized stress classification and assessment techniques can be employed by psychiatrists to aid in quantitative objective assessment of patients who undergo evaluation. These systems can also be used in the applications of talking condition intelligent automated systems in call-centers. It is very important for call-centers to take note of customers' disputes using talking condition intelligent automated systems and successfully respond to these disputes to obtain the customers' satisfaction.

It is well known that the recognition performance of speech recognition and speaker recognition systems is almost perfect under the neutral talking condition. However, the performance is degraded sharply under the shouted talking condition. Many publications show that the performance of speech recognition and speaker recognition systems under this talking condition is deteriorated significantly [4], [5], [6], [7], [8].

Speaker identification performance under the shouted talking condition is very low based on HMMs [5], [7], [8]. In previous work, Shahin focused on enhancing the recognition performance of text-dependent speaker identification systems under the shouted talking condition based on each of second-order hidden Markov models (HMM2s) and CHMM2s [7], [8].

Our work in this research differs from the work in [8] is that our work in this research focuses on enhancing the recognition performance of text-dependent speaker identification in the shouted environment based on SPHMMs using each of our collected speech database and SUSAS database. On the other hand, the work in [8] focused on enhancing speaker identification performance in the same environment based on CHMM2s. We can claim that this is the first time to use SPHMMs for speaker identification in such an environment.

This paper is organized as follows. Section 3 overviews: Hidden Markov Models (HMMs), First-Order Hidden Markov Models (HMM1s), Second-Order Hidden Markov Models (HMM2s), Circular Hidden Markov Models (CHMMs), and Second-Order Circular Hidden Markov Models (CHMM2s). Section 4 discusses the details of SPHMMs. Section 5 describes the collected speech database used. The algorithm of speaker identification based on each of CHMM2s and SPHMMs is given in Section 6. Section 7 discusses the results that are obtained in this work. Concluding remarks are drawn in Section 8.

## 3. Overview of HMMs, HMM1s, HMM2s, CHMMs, and CHMM2s

### 3.1. Hidden Markov Models

The use of HMMs in the fields of speech recognition, speaker recognition, and emotion recognition has become popular in the last three decades. HMMs have become one of the most successful and broadly used modeling techniques in the three fields [3], [4], [5], [9], [10], [11], [12], [13].

Bou-Ghazale and Hansen used HMMs in the study of evaluating the effectiveness of traditional features in recognition of speech under stress and formulating new features which are shown to enhance stressed speech recognition [3]. Zhou *et al.* applied HMMs in the study of nonlinear feature based classification of speech under stress [4]. Chen studied talker-stress-induced intraword variability and an algorithm that compensates for the systematic changes observed based on HMMs trained by speech tokens in various talking styles [5]. Nwe *et al.* exploited HMMs in the text independent method of emotion classification of speech [11]. Ververidis and Kotropoulos made use of HMMs in the classification techniques that classify speech into emotional states [12]. Bosch used HMMs to recognize emotion from the speech signal, from the viewpoint of automatic speech recognition (ASR) [13].

HMMs use Markov chain to model the changing statistical characteristics that exist in the actual observations of speech signals. HMMs are double stochastic processes where there is an unobservable Markov chain defined by a state transition matrix, and where each state of the Markov chain is associated with either a discrete output probability distribution (discrete HMMs) or a continuous output probability density function (continuous HMMs) [9]. HMMs are powerful models in optimizing the parameters that are used in modeling speech signals. This optimization decreases the computational complexity in the decoding procedure and improves the recognition accuracy [9]. This topic is widely covered in many references [9], [10].

### 3.2. First-Order Hidden Markov Models

HMM1s have been used in the training and testing phases of the vast majority of the work in the areas of speech recognition, speaker recognition, and emotion recognition [4], [5], [11], [12], [13]. The recognition performance of speech and speaker recognition systems based on HMM1s is high under the neutral talking condition; however, the performance is degraded sharply under the shouted talking condition [5], [7], [8].

In HMM1s, the underlying state sequence is a first-order Markov chain where the stochastic process is specified by a 2-D matrix of a priori transition probabilities ($a_{ij}$) between states $s_i$ and $s_j$ where $a_{ij}$ are given as [9],

$$a_{ij} = \text{Prob}(q_t = s_j | q_{t-1} = s_i) \qquad (1)$$

More details about HMM1s can be found in many references [9], [10].

### 3.3. Second-Order Hidden Markov Models

New models called HMM2s were introduced and implemented under the neutral talking condition by Mari *et al.* [14]. These models have shown to enhance the performance of speaker identification systems under the shouted talking condition [7].

The underlying state sequence in HMM2s is a second-order Markov chain where the stochastic process is specified by a 3-D matrix ($a_{ijk}$). Therefore, the transition probabilities in HMM2s are given as [14],

$$a_{ijk} = \text{Prob}\left(q_t = s_k \mid q_{t-1} = s_j, q_{t-2} = s_i\right) \tag{2}$$

with the constraints,

$$\sum_{k=1}^{N} a_{ijk} = 1 \qquad N \geq i, j \geq 1$$

The reader can obtain more details about HMM2s from references [7], [14].

### 3.4. Circular Hidden Markov Models

Most of the work performed in the fields of speech recognition, speaker recognition, and emotion recognition using HMMs has been done using Left-to-Right Hidden Markov Models (LTRHMMs) [5], [11], [12], [13]. LTRHMMs yield high speaker identification performance under the neutral talking condition; however, the performance is deteriorated sharply under the shouted talking condition [8].

CHMMs were introduced and used by Zheng and Yuan under the neutral talking condition [15]. These models yield high speaker identification performance under the neutral talking condition [16]. CHMMs have the following properties:

1. The underlying Markov chain has no final or absorbing state. Therefore, the corresponding HMMs can be trained by as long training sequence as desired.

2. Once the Markov chain leaves any state, that state can be revisited only at the next time.

In CHMMs, the state transition coefficients have the property of [8],

$$a_{ij} = a_{ji} \qquad i, j = 1,2,\ldots,N \qquad (3)$$

Therefore, the state transition probability matrix *A* possesses symmetry. More information about CHMMs can be found in [15], [16].

### 3.5. Second-Order Circular Hidden Markov Models

New models called CHMM2s have been proposed and implemented by Shahin to enhance the performance of text-dependent speaker identification under the shouted talking condition [8]. CHMM2s possess the characteristics of both CHMMs and HMM2s:

1. The underlying state sequence in HMM2s is a second-order Markov chain where the stochastic process is specified by a 3-D matrix because in these models the state-transition probability at time *t*+1 depends on the states of the Markov chain at two times *t* and *t*-1. On the other hand, the underlying state sequence in HMM1s is a first-order Markov chain where the stochastic process is specified by a 2-D matrix because in these models it is assumed that the state-transition probability at time *t*+1 depends only on the state of the Markov chain at one time *t*. Hence, the stochastic process that is specified by a 3-D matrix gives better speaker identification performance than that specified by a 2-D matrix [7].

2. The Markov chain in CHMMs is more powerful in modeling the changing statistical characteristics that exist in the actual observations of speech signals than that in LTRHMMs.

3. In LTRHMMs, the absorbing state governs the fact that the rest of a single observation sequence provides no further information about earlier states once the underlying Markov chain reaches the absorbing state. In speaker identification, it is true that the Markov chain should be able to revisit the earlier states because the states of HMMs reflect the vocal organic configuration of the speaker. Therefore, the vocal organic configuration of the speaker is reflected to states more conveniently using CHMMs than using LTRHMMs.

In the training phase of CHMM2s, the initial elements of the parameters are chosen to be [8],

$$v_k(i) = \frac{1}{N} \qquad\qquad N \geq i, k \geq 1 \qquad(4)$$

where, $v_k(i)$ is the initial element of the probability of an initial state distribution and $N$ is the number of states.

$$\alpha_1(i,k) = v_k(i) b_{ki}(O_1) \qquad\qquad N \geq i, k \geq 1 \qquad(5)$$

where, $\alpha_1(i,k)$ is the initial element of the forward probability of producing the observation vector $O_1$.

$$a^1_{ijk} = \begin{cases} \frac{1}{3} & i = 1, j, k = 1, 2, ..., N \\ \frac{1}{3} & N-1 \geq i \geq 2, i+1 \geq j \geq i-1, N \geq k \geq 1 \\ \frac{1}{3} & i = N, j, k = 1, 2, ..., N \\ 0 & \text{otherwise} \end{cases} \qquad(6)$$

where, $a^1_{ijk}$ is the initial element of $a_{ijk}$. $a_{ijk}$ is the state transition coefficient from state $S_i$ to state $S_k$.

$$b^1_{ijk} = \frac{1}{M} \qquad\qquad N \geq j, k \geq 1, M \geq i \geq 1 \qquad(7)$$

where, $b^1_{ijk}$ is the initial element of the observation symbol probability and $M$ is the number of observation symbols.

$$\beta_T(j,k) = \frac{1}{N} \qquad\qquad N \geq j, k \geq 1 \qquad(8)$$

where $\beta_T(j,k)$ is the initial element of the backward probability of producing the observation vector $O_T$.

The probability of the observation vector $O$ given the CHMM2s model $\Phi$, can be calculated by,

$$P(O|\Phi) = \sum_{k=1}^{N} \sum_{i=1}^{N} \alpha_T(i,k) \qquad (9)$$

## 4. Suprasegmental Hidden Markov Models

A suprasegmental is a vocal effect that extends over more than one sound segment in an utterance, such as pitch, stress, or juncture pattern. Suprasegmental is often used for tone, vowel length, and features like nasalization and aspiration.

SPHMMs possess the ability and the capability to summarize several states of HMMs into what is called a suprasegmental state. Suprasegmental state can look at the observation sequence through a larger window. Such a state allows observations at rates appropriate for the situation of modeling. For example, prosodic information can not be observed at a rate that is used for acoustic modeling. The fundamental frequency, intensity, and duration of speech signals are the main acoustic parameters that describe prosody [17]. The prosodic features of a unit of speech are called suprasegmental features because they affect all the segments of the unit. Therefore, prosodic events at the levels of: phone, syllable, word, and utterance are modeled using suprasegmental states; on the other hand, acoustic events are modeled using conventional hidden Markov states.

Prosodic information can be combined and integrated with acoustic information within HMMs [18]. This combination and integration can be performed as given by the following formula,

$$log\ P(\lambda^v, \Psi^v | O) = (1-\alpha).\ log\ P(\lambda^v | O) + \alpha.\ log\ P(\Psi^v | O) \qquad (10)$$

where,

$\alpha$ : is a weighting factor that is chosen to be equal to 0.5 (so no biasing towards any model).

$\lambda^v$: is the acoustic model for the $v$th speaker.

$\Psi^v$: is the suprasegmental model for the $v$th speaker.

$O$: is the observation vector or sequence of an utterance.

Equation 10 demonstrates that leaving a suprasegmental requires adding the log probability of this suprasegmental state given the respective suprasegmental observations within the speech signal to the log probability of the current acoustic model given the respective acoustic observations within the speech signal.

$P(\lambda^v | O)$ and $P(\Psi^v | O)$ can be calculated using Bayes theorem as given in equation 11 and equation 12 respectively [10],

$$P(\lambda^v | O) = \frac{P(O | \lambda^v) P_0(\lambda^v)}{P(O)} \qquad (11)$$

$$P(\Psi^v | O) = \frac{P(O | \Psi^v) P_0(\Psi^v)}{P(O)} \quad (12)$$

where, $P_0(\lambda^v)$ and $P_0(\Psi^v)$ are the priori distribution of the acoustic model and the suprasegmental model respectively. The parameter priori distribution characterizes the statistics of the parameters of interest before any measurement is made.

Yegnanarayana *et al.* used suprasegmental and source features, besides spectral features in their proposed method for text-dependent speaker verification systems. The combination of suprasegmental, source, and spectral features significantly enhances the performance of speaker verification system [19].

In the last three decades, the majority of the work performed in the fields of speech recognition and speaker recognition on HMMs has been done using LTRHMMs because phonemes follow strictly the left to right (LTR) sequence [5], [11], [12], [13]. In this work, LTRSPHHMs is obtained and derived from LTRHMMs. Therefore, the relationship between HMMs states and SPHMMs states is codified (HMMs and SPHMMs are evolved dependently). Fig. 1 shows an example of a basic structure of LTRSPHMMs that is derived from LTRHMMs. In this figure, $q_1$, $q_2$, ..., $q_6$ are hidden Markov states. $p_1$ is a suprasegmental state (e.g. phone) that consists of $q_1$, $q_2$, and $q_3$. $p_2$ is a suprasegmental state (e.g. phone) that is made up of $q_4$, $q_5$, and $q_6$. $p_3$ is a suprasegmental state (e.g. syllable) that is composed of $p_1$ and $p_2$. $a_{ij}$ is the transition probability between the *i*th hidden

Markov state and the *j*th hidden Markov state, while $b_{ij}$ is the transition probability between the *i*th suprasegmental state and the *j*th suprasegmental state.

Since the stressful cues contained in an utterance can not be assumed as specific sequential events in the signal, an ergodic or fully connected HMMs structure becomes more appropriate than LTR structure because every state in the ergodic structure can be reached in a single step from every other state. An ergodic or fully connected SPHMMs that are derived from ergodic or fully connected HMMs have been used in this work.

## 5. Speech Database

Our speech database was collected from 30 (15 males and 15 females) healthy adult Native American speakers. Each speaker uttered 8 sentences where each sentence was uttered 5 times in one session (training session) and 4 times in another separate session (testing session) under each of the neutral and shouted talking conditions. These sentences were:

1) *He works five days a week.*
2) *The sun is shining.*
3) *The weather is fair.*
4) *The students study hard.*
5) *Assistant professors are looking for promotion.*
6) *University of Sharjah.*
7) *Electrical and Computer Engineering Department.*
8) *He has two sons and two daughters.*

Our speech database was captured by a speech acquisition board using a 12-bit linear coding A/D converter and sampled at a sampling rate of 12 kHz. Our database was a 12-bit per sample linear data. Each speech signal under each of the neutral and shouted talking conditions was applied every 5 ms to a 30 ms

Hamming window. 12*th* order linear prediction coefficients (LPCs) were extracted from each frame by the autocorrelation method. The 12*th* order LPCs were then transformed into 12*th* order linear prediction cepstral coefficients (LPCCs). The LPCC feature analysis was used to form the observation vectors in each of CHMM2s and SPHMMs.

In CHMM2s, the number of states, *N*, was 9. The number of mixture components, *M*, was 5 per state, with a continuous mixture observation density was selected for these models. In ergodic SPHMMs, the number of suprasegmental states was 3 ($p_1$, $p_2$, and $p_3$). $p_1$ was composed of 3 hidden Markov states: $q_1$, $q_2$, and $q_3$, $p_2$ was made up of 3 hidden Markov states: $q_4$, $q_5$, and $q_6$, while $p_3$ consisted of 3 hidden Markov states: $q_7$, $q_8$, and $q_9$. Therefore, each three states of HMMs in this research are summarized into one suprasegmental state. The number of mixture components was 10 per state, with a continuous mixture observation density was selected for ergodic SPHMMs. Fig. 2 shows our adopted 3-state ergodic suprasegmental hidden Markov model. The transition matrix, *A*, of this structure can be written in terms of the $b_{ij}$ coefficients (positive coefficients) as,

$$A = \begin{bmatrix} b_{11} & b_{12} & b_{13} \\ b_{21} & b_{22} & b_{23} \\ b_{31} & b_{32} & b_{33} \end{bmatrix}$$

Our database in each of CHMM2s and SPHMMs was divided into training data under the neutral talking condition and test data under each of the neutral and shouted talking conditions. Our speech database in this work was a "closed set".

## 6. The Algorithm of Speaker Identification Based on Each of CHMM2s and SPHMMs

### 6.1 The Algorithm Based on CHMM2s

In the training session of CHMM2s, a reference model was derived using 5 of the 9 utterances per the same speaker per the same sentence under the neutral talking condition. Therefore, each speaker per sentence under the neutral talking condition was represented by one reference model. This session was composed of 1,200 utterance.

In the testing (identification) session (completely separate from the training session), each one of the 30 speakers used 4 of the 9 utterances per the same sentence (text-dependent) under each of the neutral and shouted talking conditions. This session consisted of 1,920 utterance. The probability of generating every utterance was computed (there were 30 probabilities per utterance for each talking condition), the model with the highest probability was chosen as the output of speaker identification as given in the following formula,

$$V^* = \arg\max_{30 \geq v \geq 1} \left\{ P\left(O|\Phi^v\right) \right\} \qquad (13)$$

where $V^*$ is the index of the identified speaker, $O$ is the observation vector or sequence that belongs to the unknown speaker, and $P(O|\Phi^v)$ is the probability of the observation sequence $O$ given the $v$th CHMM2s model $\Phi^v$. A block diagram of speaker identification based on CHMM2s is shown in Figure 3.

## 6.2 The Algorithm Based on SPHMMs

The training session of SPHMMs was very similar to the training phase of the conventional HMMs. In the training phase of SPHMMs, suprasegmental models were trained on top of acoustic models of HMMs. In this phase, a reference model was derived using 5 of the 9 utterances per the same speaker per the same sentence under the neutral talking condition. Hence, each speaker per sentence was represented by one reference model under the neutral talking condition.

In the testing session (completely separate from the training session), each one of the 30 speakers used 4 of the 9 utterances per the same sentence under each of the neutral and shouted talking conditions. The probability of generating every utterance was computed, the model with the highest probability was chosen as the output of speaker identification as given in the following formula,

$$V^* = \arg\max_{30 \geq v \geq 1} \left\{ P\left(O \mid \lambda^v, \Psi^v \right) \right\} \qquad (14)$$

where, $O$ is the observation vector or sequence that belongs to the unknown speaker. A block diagram of speaker identification based on SPHMMs is shown in Figure 4.

## 7. Results and Discussion

Our work in this research focused on using SPHMMs in the training and testing phases of text-dependent speaker identification in the shouted environment. This is the first known investigation into SPHMMs evaluated in such an environment for speaker identification.

Table 1 and Table 2 summarize the results of speaker identification performance under each of the neutral and shouted talking conditions using our collected speech database based on SPHMMs and CHMM2s respectively. We compare our results obtained based on SPHMMs with that obtained based on CHMM2s. The average improvement rate of using SPHMMs over CHMM2s under the shouted talking condition is significant (10.3%). Therefore, SPHMMs are superior models over CHMM2s under such a talking condition. This may be attributed to the following reasons:

1) SPHMMs are suitable models to integrate observations from the shouted modality because such models allow for observation at a rate appropriate for the shouted modality.

2) Suprasegmental states are used to capture the prosodic properties of words and utterances because they allow us to make observations at a time scale suitable for prosodic phenomena.

On the other hand, CHMM2s are not efficient models as SPHMMs to integrate observations from the shouted modality because observations within the conventional CHMM2s have to be at a constant rate.

In this work, speaker identification performance under the neutral talking condition has been slightly improved based on SPHMMs compared to that based on CHMM2s. Our results show that speaker identification performance has been improved from 98% based on CHMM2s to 99% based on SPHMMs. The average improvement rate of using SPHMMs over CHMM2s under such a talking condition is 1.02%. This insignificant improvement resulted from the fact that CHMM2s have proven to be powerful and efficient models under the neutral talking condition [8].

To verify that SPHMMs yield better speaker identification performance than CHMM2s in the shouted environment, we use SUSAS database. Since this database does not contain utterances under the shouted talking condition, we decided to use the angry talking condition as an alternative talking condition to the shouted talking condition (since the shouted talking condition can not be entirely separated from the angry talking condition in real life). SUSAS database has been used in the training and testing phases of speaker identification under each of the neutral and angry talking conditions (part of this database consists of 9 male speakers uttering speech signals under each of the two talking conditions). Table 3 summarizes the results of speaker identification performance based on each of SPHMMs and CHMM2s under each of the neutral and angry talking conditions using SUSAS database. It is evident from Table 3 that SPHMMs significantly enhance speaker identification performance under the angry talking condition compared to that using CHMM2s. The improvement rate is 11.3%.

## 8. Concluding Remarks

The performance of speaker identification in the shouted environment is degraded sharply compared to that in the neutral environment. Our work shows that SPHMMs are superior models over CHMM2s in the shouted environment using each of our collected speech database and SUSAS database. However, the improvement based on SPHMMs is limited. A thorough study and work using new algorithms and models are needed to achieve better speaker identification improvement in the shouted environment.

There are some limitations in our work. First, the number of speakers in SUSAS database is limited to 9. Second, all the 9 speakers in SUSAS database are of the same gender (male). Finally, the angry talking condition in SUSAS database is used as an alternative talking condition to the shouted talking condition.

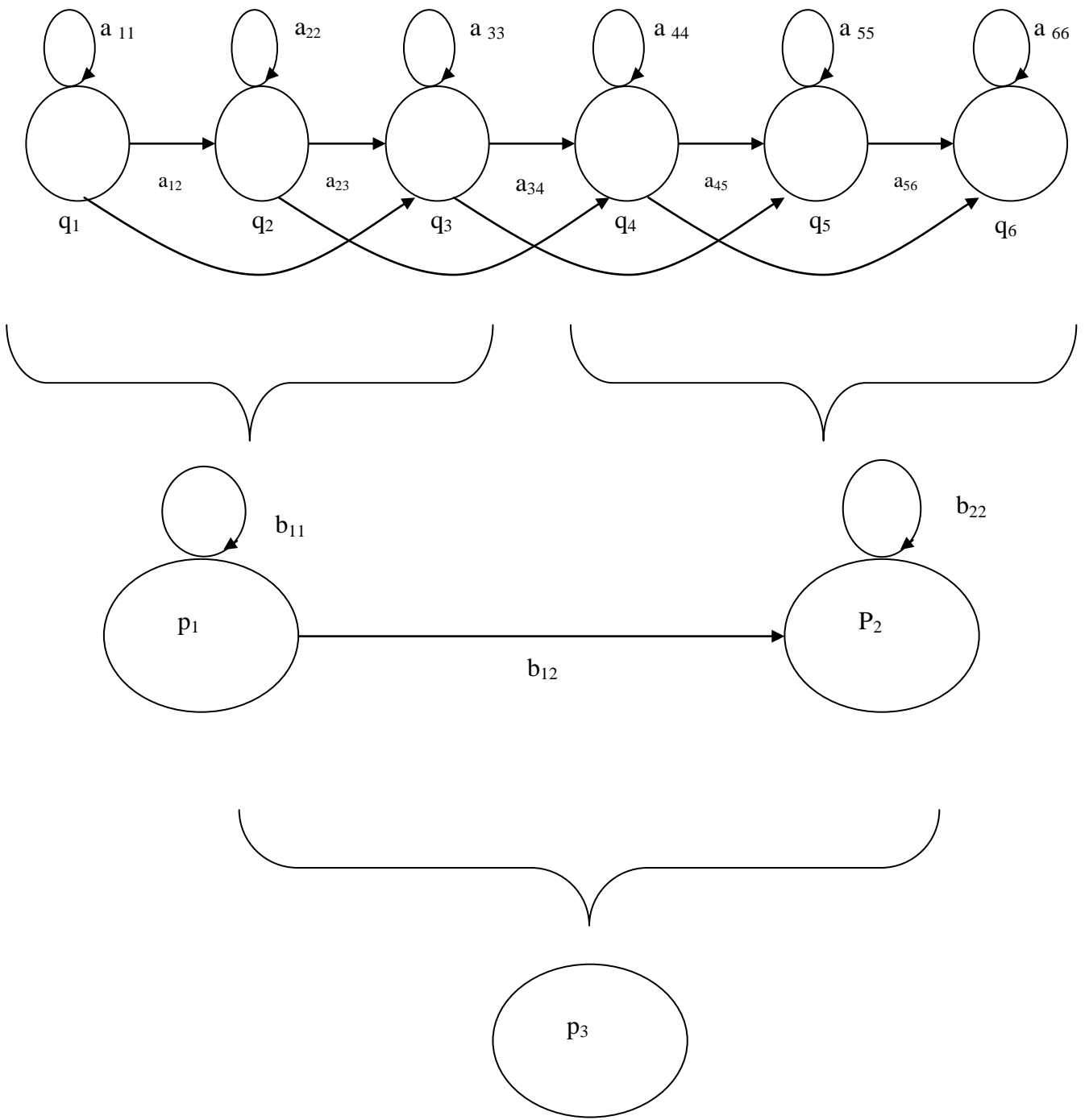

**Figure 1.** Basic structure of LTRSPHMMs derived from LTRHMMs

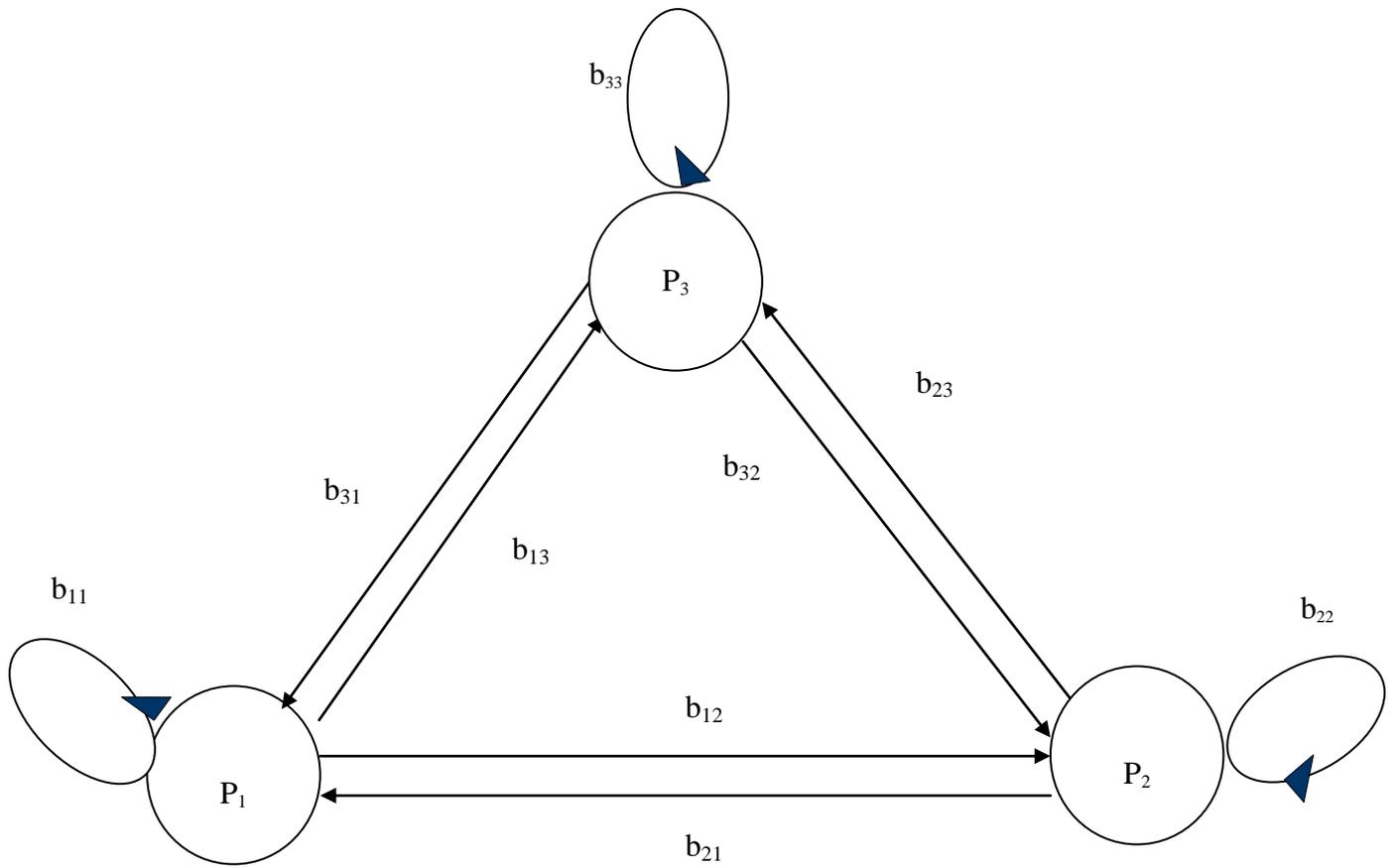

**Figure 2.** 3-state ergodic SPHMMs

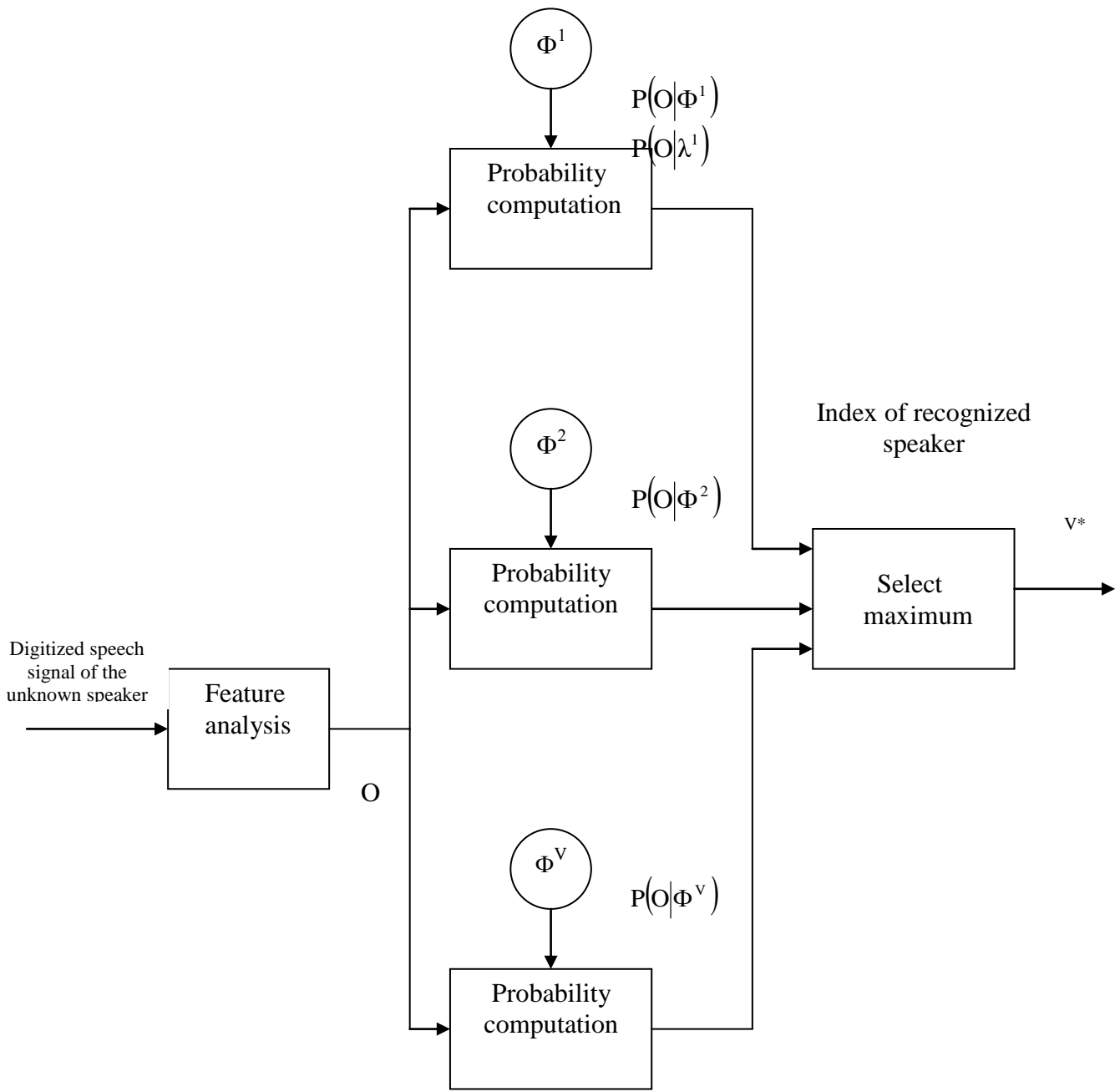

**Figure 3.** Block diagram of speaker identification based on CHMM2s

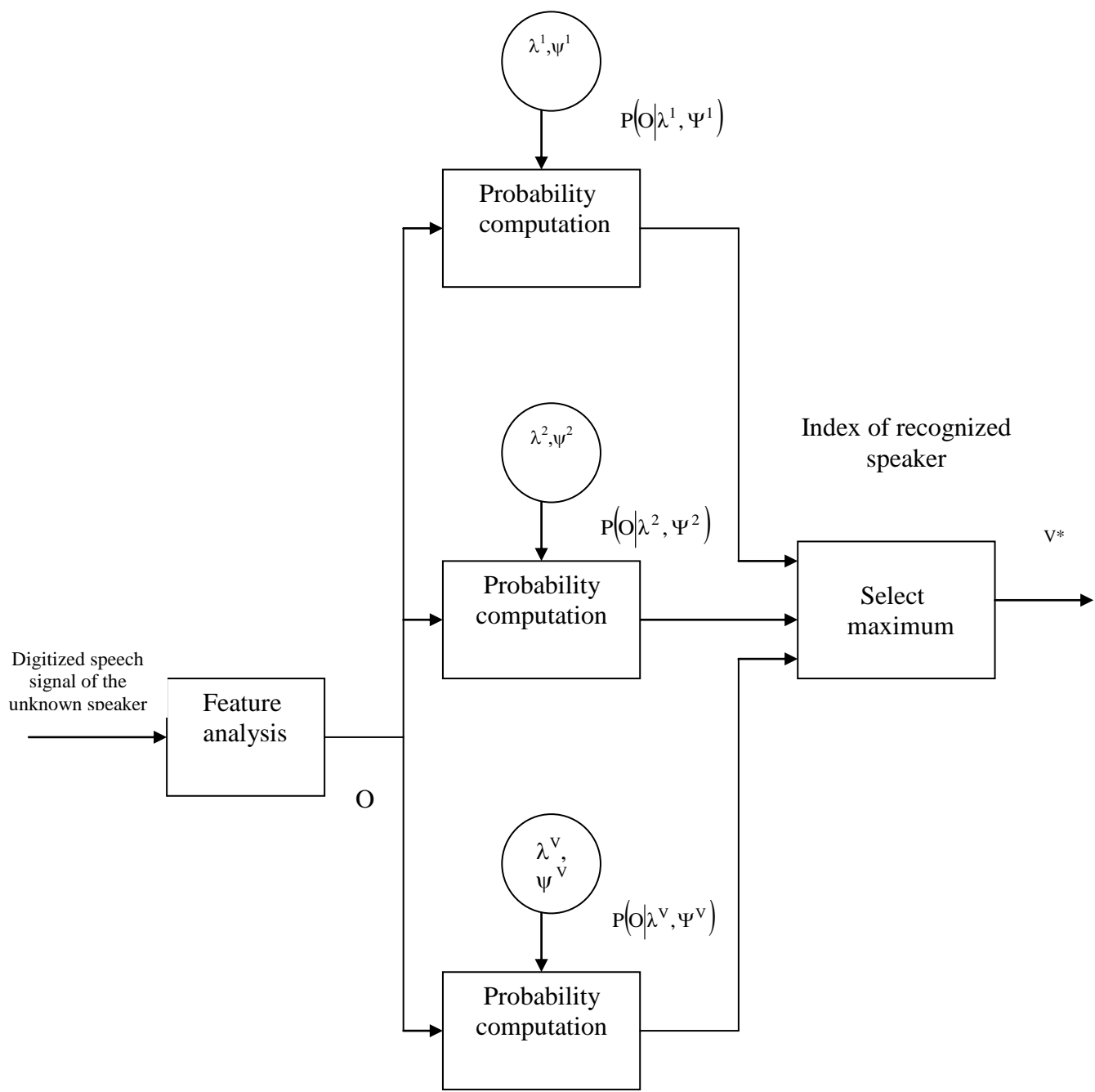

**Figure 4.** Block diagram of speaker identification based on SPHMMs

Table 1

Speaker identification performance under each of the neutral and shouted talking conditions based on SPHMMs using our collected speech database

| Gender | Neutral talking condition (%) | Shouted talking condition (%) |
|---|---|---|
| Male | 99 | 74 |
| Female | 99 | 76 |
| Average | 99 | 75 |

Table 2

Speaker identification performance under each of the neutral and shouted talking conditions based on CHMM2s using our collected speech database

| Gender | Neutral talking condition (%) | Shouted talking condition (%) |
|---|---|---|
| Male | 98 | 68 |
| Female | 98 | 68 |
| Average | 98 | 68 |

Table 3

Speaker identification performance based on each of SPHMMs and CHMM2s under each of the neutral and angry talking conditions using SUSAS database

| Models | Neutral talking condition (%) | Angry talking condition (%) |
|---|---|---|
| SPHMMs | 99 | 79 |
| CHMM2s | 99 | 71 |